\documentclass[twocolumn]{IEEEtran} 
       \usepackage{epsfig}%

\def\BibTeX{{\rm B\kern-.05em{\sc i\kern-.025em b}\kern-.08em
    T\kern-.1667em\lower.7ex\hbox{E}\kern-.125emX}}

\begin{document}

\title{Prototype tests for the ALICE TRD} 

\author{A.~Andronic, H.~Appelsh\"auser, C.~Blume, 
P.~Braun-Munzinger, V.~C\u at\u anescu, M.~Ciobanu,
H.~Daues, A.~Devismes, Ch.~Finck, N.~Herrmann,
T.~Lister, T.~Mahmoud, T.~Peitzmann, M.~Petrovici, 
A.~Reischl, K.~Reygers, R.~Santo, R.~Schulze, S.~Sedykh,
R.S.~Simon, J.~Stachel, H.~Stelzer, J.~Wessels, 
O.~Winkelmann, B.~Windelband, C.~Xu \\ 
(for the ALICE Collaboration) \\
\thanks{Manuscript received November 5, 2000.}
\thanks{A.~Andronic, C.~Blume, P.~Braun-Munzinger, H.~Daues, 
A.~Devismes, Ch.~Finck, R.~Schulze, S.~Sedykh, R.S.~Simon and H.~Stelzer
are with Gesellschaft f\"ur Schwerionenforschung, D-64291 Darmstadt, Germany.
(Correspondence to: Anton Andronic, GSI, Planckstr. 1, D-64291 Darmstadt, 
telephone: +40-6159-712769, e-mail: andronic@gsi.de).
}
\thanks{H.~Appelsh\"auser, N.~Herrmann, T.~Mahmoud, A.~Reischl, J.~Stachel,
J.~Wessels, B.~Windelband and C.~Xu are with 
Physikalisches Institut der Universit\"at Heidelberg,
D-69120 Heidelberg, Germany.}
\thanks{V.~C\u at\u anescu, M.~Ciobanu and M.~Petrovici are with
NIPNE Bucharest, 76900 Bucharest-Magurele, Romania.}
\thanks{T.~Lister, T.~Peitzmann, K.~Reygers, R.~Santo and  O.~Winkelmann
are with Institut f\"ur Kernphysik, Universit\"at M\" unster, 
D-48149 M\"unster, Germany.}
}

\markboth{IEEE Transactions On Nuclear Science} {TRD} 

\maketitle

\begin{abstract}
A Transition Radiation Detector (TRD) has been designed to improve the 
electron identification and trigger capability of the ALICE experiment 
at the Large Hadron Collider (LHC) at CERN. 
We present results from tests of a prototype of the TRD concerning 
pion rejection for different methods of analysis over a momentum range 
from 0.7 to 2 GeV/c.
We investigate the performance of different radiator types, composed of 
foils, fibres and foams.
\end{abstract}

\begin{keywords}
Transition Radiation Detector, radiator performance, electron identification,
pion rejection.
\end{keywords}

\section{Introduction}
\PARstart{T}{ransition} Radiation Detectors (TRDs) are  
being used in high energy experiments to improve the identification of
electrons with respect to pions for momenta between $\sim$1 and 100 GeV/c
(see \cite{dol} for a review on TRDs). 
A proposal to add a TRD \cite{trd} to the ALICE experiment \cite{alice} was 
approved in May 1999.
By increasing the pion rejection power by at least a factor of 100 for momenta 
above 2~GeV/c, the TRD will allow, in conjunction with other ALICE detectors, 
to study, in the central region of the ALICE detector, various aspects of 
dielectron physics, among them the production of quarkonia like $J/\psi$, 
$\psi'$ and the members of the $\Upsilon$ family,
as well as the production of open charm and open beauty \cite{trd}.

\section{Detector characteristics}
The ALICE TRD is composed of a radiator and a photon detector, 
the latter being a Drift Chamber (DC) with a 3~cm drift zone and an 
amplification region of 6~mm.
To cope with the large charged particles multiplicities expected in Pb+Pb 
collisions at LHC
and to provide the necessary position resolution for track reconstruction,
the readout of the DC is done on a chevron pad plane \cite{yu}.
We use the chevron width of 10~mm and the chevron step is tailored to the 
anode wire pitch of 5~mm. Each pad is connected to a preamplifier (PA) whose 
output is fed into a Flash-ADC (FADC), sampling with a frequency of about 
20 MHz the drift time of up to 2~$\mu$s.
The detection gas of the DC will be a Xe-based mixture to facilitate an 
efficient absorption of the transition radiation (TR) photons with typical 
energies between 4 and 30 keV.
Six layers will surround the interaction point in full azimuth at radial 
distances from 2.9 to 3.7 meters and will match in polar angle the acceptance 
of the Time Projection Chamber ($45^\circ\le\theta\le135^\circ$).
A total number of 540 subdetectors will add up to a total surface of the TRD 
of about 800~m$^2$, with the largest single module of 1.22$\times$1.56~m$^2$.
With a pad size of 4.5-6~cm$^2$, the total number of channels will be up to 
1.2 million, depending on the final geometrical configuration.

\section{Results of prototype tests}
A prototype of a DC with the characteristics presented above has been 
built at GSI. Apart from the mechanical realization, it differs from a final 
detector also concerning the surface, which is 0.5$\times$0.6~m$^2$. 
We use W-Au anode wires of 20~$\mu$m diameter and Cu-Be cathode 
wires of 75~$\mu$m diameter. The entrance window of 25~$\mu$m aluminized 
kapton serves simultaneously as gas barrier and drift electrode.

\begin{figure}[hbt]
\centering\mbox{\epsfig{file=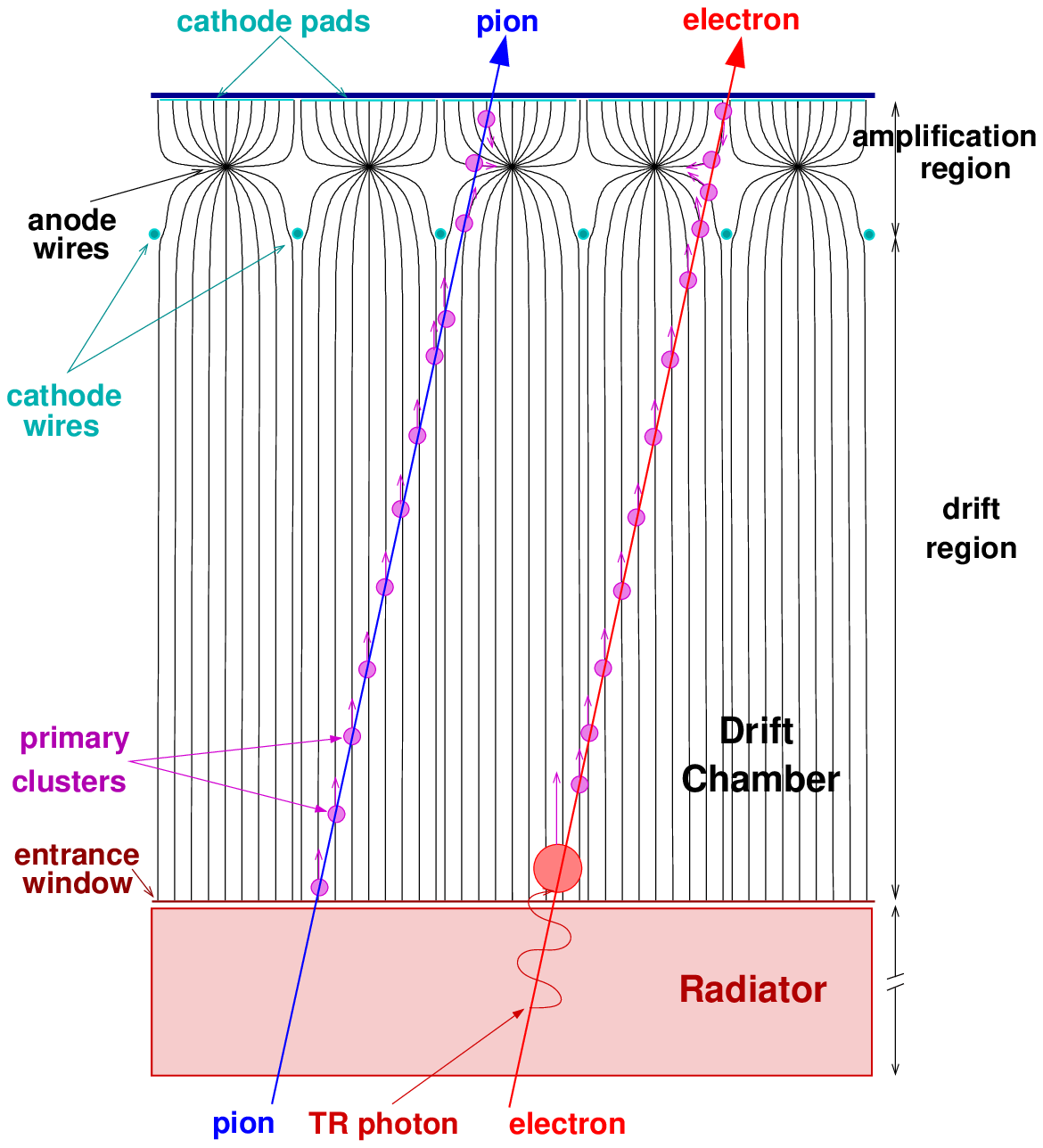, width=.48\textwidth}}
\caption{The geometry of one TRD module. The geometric proportions and 
the field lines in the Drift Chamber (DC) are accurate. 
The radiator is not to scale.
Schematic signals produced by a pion and an electron are shown.}
\label{fig-1} \end{figure} 

A cross-section of a segment of one module of the TRD is shown in 
Fig.~\ref{fig-1} along with a schematic illustration of the signals 
detected by the DC from a pion and an electron (with Lorentz factor 
$\gamma>$1000).
The signal from TR photons produced in the radiator is superimposed to 
the signal from energy loss of electrons, as their angle with respect 
to the electron trajectory is very small (of the order of 1/$\gamma$).
The field lines in the DC, depicted in Fig.~\ref{fig-1}, are calculated 
with GARFIELD \cite{rob} for the dimensions mentioned above.
The radiator is not to scale.

The prototype has been tested with Ar- and Xe-based gas mixtures, using a 
$^{55}$Fe X-ray source of 5.9~keV, cosmic rays and mixed electron-pion
beams provided by the secondary pion beam facility at GSI.
Current- and charge-sensitive PAs were specially designed and built for 
these tests. For the results presented in the following the charge-sensitive 
PA was used. It has a gain of 2 mV/fC and a noise of about 2000 electrons rms.
We use an 8-bit non-linear FADC system with 100 MHz sampling, integrated 
in the GSI-standard, VME-based data acquisition system, MBS \cite{mbs}.

\subsection{Source tests}

The $^{55}$Fe source was used to determine for the DC the optimum voltages 
and for measuring the gas gain and the energy resolution.
By integrating the FADC pulse on three adjacent pads to account for the 
charge sharing, the energy resolution at the full energy peak of 5.9~keV 
is about 10\%.

\subsection{Beam tests}
\begin{figure}[hbt]
\vspace{3mm}
\centering\mbox{\epsfig{file=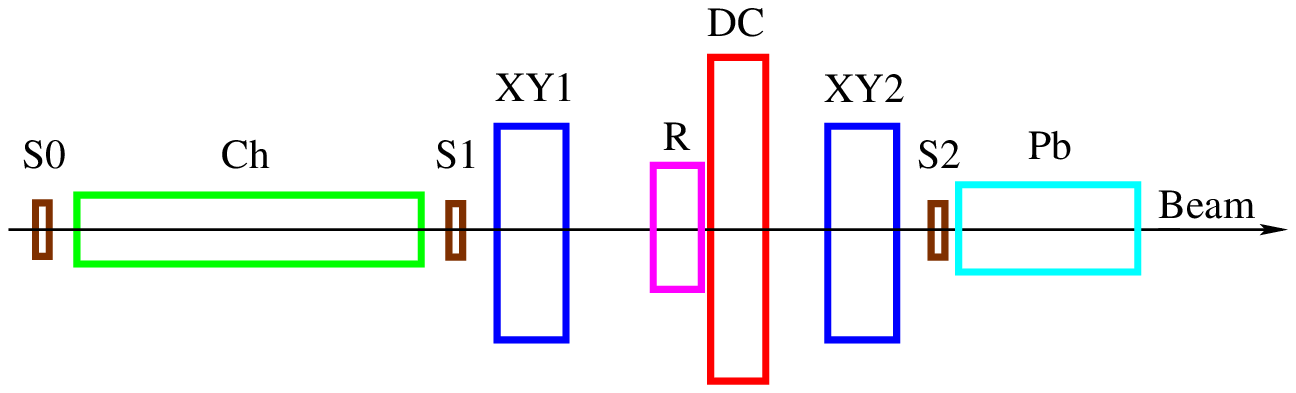, width=.49\textwidth}}
\caption{The setup used for the beam tests (not to scale). The different 
components are explained in the text.}
\label{fig-2} \end{figure}

The setup used for the beam tests is sketched in Fig.~\ref{fig-2}. 
In addition to the DC and the radiator (R), the setup consisted of 
three scintillator counters (S0, S1, S2),
a gas-filled threshold Cherenkov detector (Ch), 
two position-sensitive detectors (XY1, XY2) 
- either multiwire proportional chambers or silicon strip detectors -
and a Pb-glass calorimeter (Pb).

\begin{figure}[hbt]
\centering\mbox{\epsfig{file=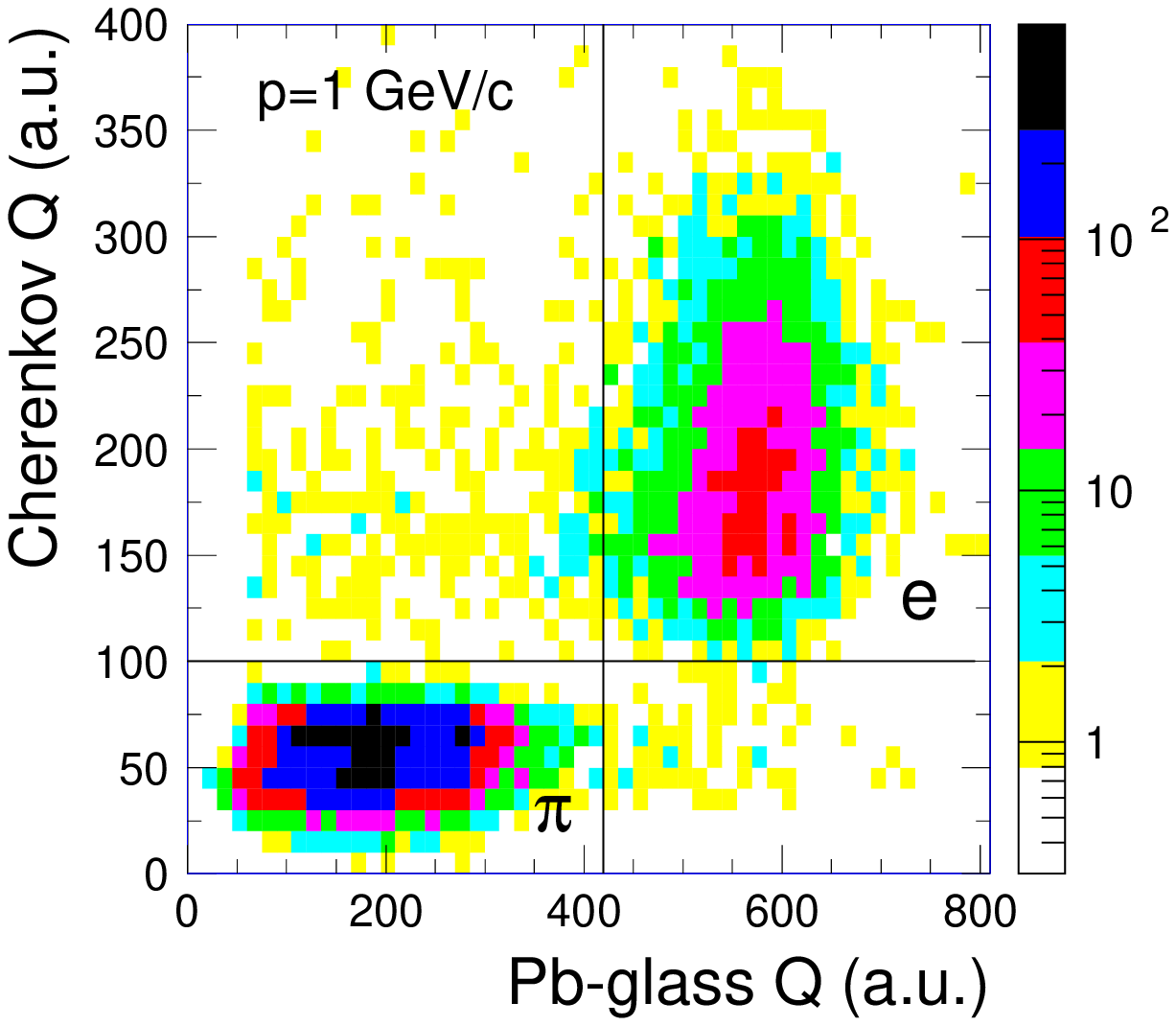, width=.49\textwidth}}
\caption{The correlation of the signals from the Cherenkov detector and
the Pb-glass calorimeter. The thresholds used to identify pions and electrons
are plotted.}
\label{fig-3} \end{figure}

The measurements have been carried out at beam momenta between 0.7 and 
2~GeV/c.
The electron content of the beam varies as function of momentum, being of the
order of 2-3\% for 1~GeV/c.
The beam trigger was defined by the S1 and S2 scintillator counters, 
to which the Cherenkov signal was added as the electron trigger. 
Both electron and pion events can be acquired during one spill by using
appropriate pion scaledown factors. 
Off-line the events were selected using the correlation between the signals
delivered by the Cherenkov and the Pb-glass detectors, shown in 
Fig.~\ref{fig-3} for the momentum of 1~GeV/c.
As seen, by requiring threshold energy deposits in both
detectors (the lines in Fig.~\ref{fig-3}) one can isolate clean samples 
of pions and electrons. For this momentum we used a pion scaledown factor of 8.

The gas mixture used for the DC was 90\%~Xe, 10\%~CH$_4$ and the voltages 
were U$_d$=-4.0 kV for the drift plane and U$_a$=1.6 kV for the
anode wires. At these voltages the gas gain of the chamber was about 8000.
The oxygen content in the gas was continuously monitored, being kept below 
10~ppm using a flow of about 2~liters/hour.

\begin{table}[htb]
\caption{The properties of various radiators.} \label{tab-1}
\begin{center}\begin{tabular}{llll}\hline
Name     & Material & $\rho$ (g/cm$^3$)   & d ($\mu$m) \\ \hline
foils-1  & PP       & 120 foils          & 20/500 \\
foils-2  & PP       & 220 foils          & 25/250 \\ 
fibres-1 & PP       & 0.09               & 17 \\
fibres-2 & PP       & 0.05               & 15-20 \\ 
foam-1   & PP       & 0.03               & 1300 \\
foam-2   & PP       & 0.06               & 700 \\
foam-3   & RC       & 0.11               & $\le$10 \\
foam-4   & RC       & 0.11               & 700 \\
foam-5   & PE       & 0.12               & 800 \\ \hline

\end{tabular}
\end{center}\end{table}

Different radiators were tested: regular foils of polypropylene (PP),
mats of irregular PP fibres with various fiber diameters 
(between 15 and 33 $\mu$m) and foams of different material type: PP, 
polyethylene (PE) and Rohacell (RC).
These radiators spanned a large range in densities and structural properties, 
as one can see in Table~\ref{tab-1}, where we present their characteristics.
The quantity $d$ quoted here is the linear dimension of the structural unit, 
which for the foils means foil/gap thicknesses, for the fibres the diameter 
and for the foams the typical pore size.
The thicknesses range is also large, from 3 to 10~cm.
Unless otherwise specified, the {\it fibres-17$\mu$m} radiator employed for the 
following plots is the {\it fibres-1} radiator with X=0.3~g/cm$^2$.

\begin{figure}[htb]
\centering\mbox{\epsfig{file=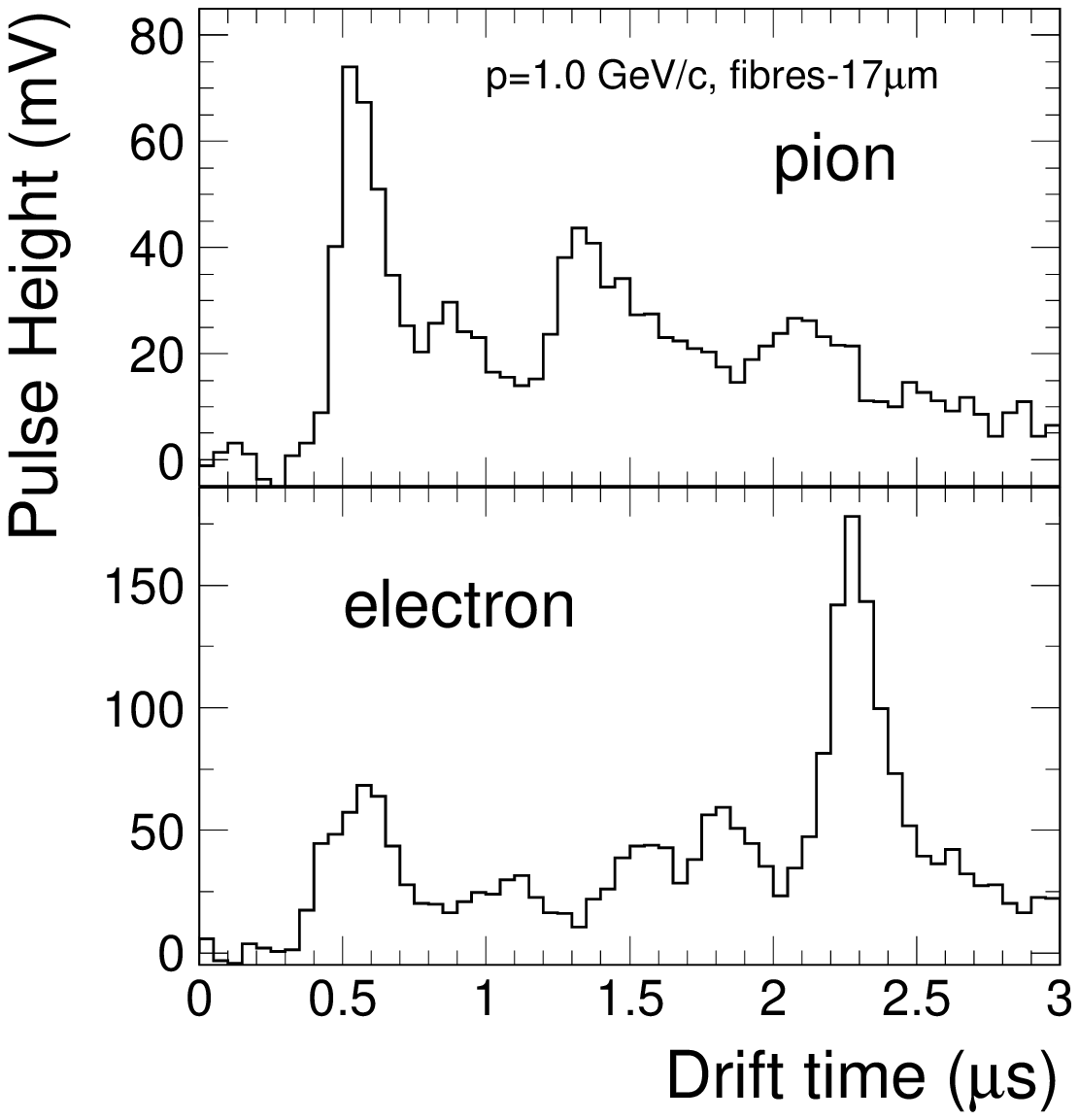, width=.44\textwidth}}
\caption{Typical signals as function of drift time for pions and electrons for
the momentum of 1.0~GeV/c. Note the different scales on the vertical axis.}
\label{fig-4} \end{figure} 

Detailed simulations \cite{trd} showed that the electron identification
is significantly improved by using, along with the pulse height, the drift 
time information.
The need for FADCs is driven in addition by the necessary tracking 
capabilities in conjunction with the trigger option for electrons with 
momenta above 3~GeV/c \cite{trd}.
In Fig.~\ref{fig-4} we show an example of the signal distribution as a 
function of the drift time for a pion and an electron. 
Here and in all the subsequent analyses we are using a time bin of 50~ns, 
a value similar to the one in the final configuration of the TRD in ALICE. 
Note the different amplitude of the two signals and, for electrons, the 
big cluster at the late drift time, corresponding possibly to a TR photon
absorbed early at the entrance of the DC (see Fig.~\ref{fig-1}). 
The time zero has been arbitrarily shifted by about 0.4~$\mu$s to
have a measurement of the baseline.

\begin{figure}[hbt]
\centering\mbox{\epsfig{file=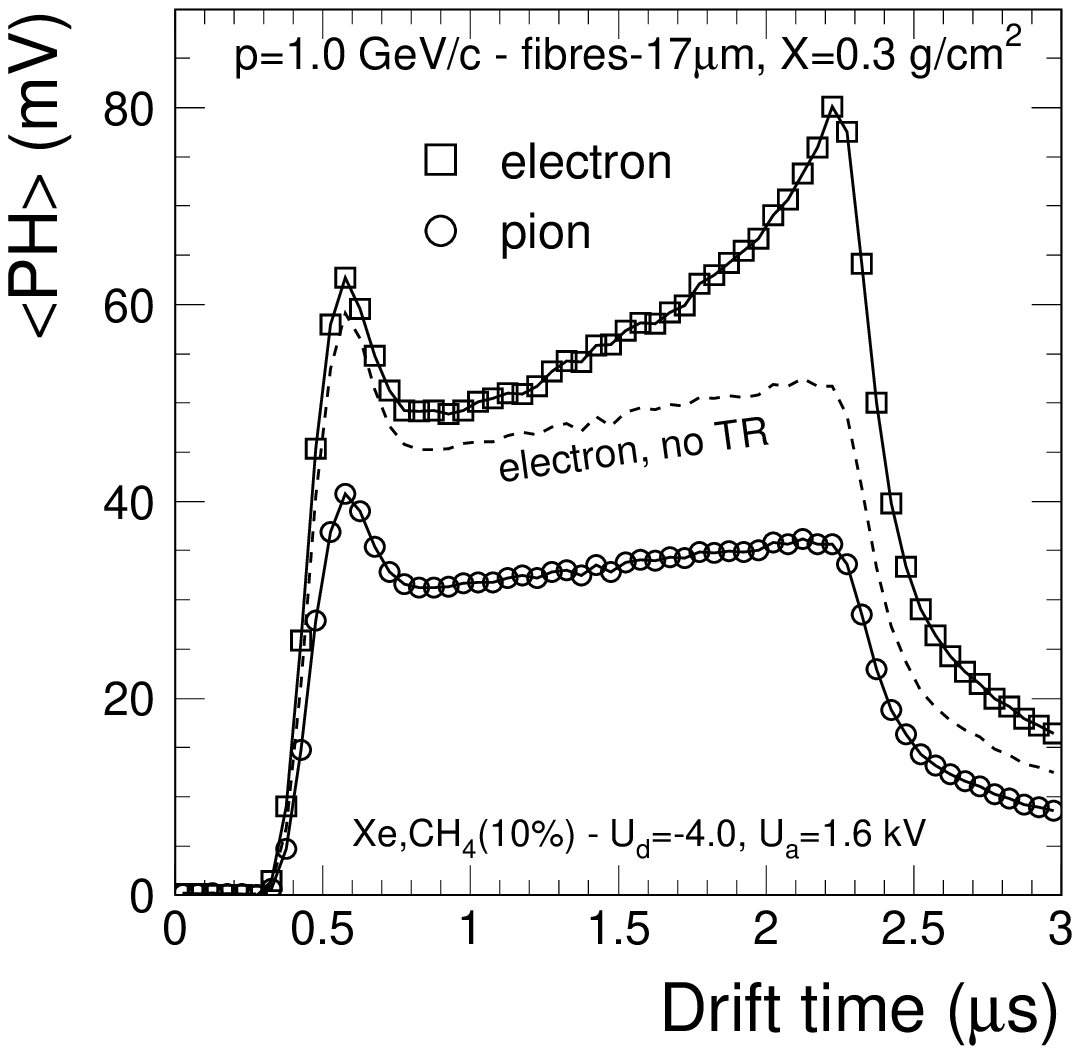, width=.44\textwidth}}
\caption{Average pulse height as function of the drift time for pions 
and electrons for a radiator of 17 $\mu$m diameter fibres at the momentum 
of 1~GeV/c.
The dashed line represents the energy deposit by pure ionization in case 
of electrons (see text).}
\label{fig-5} \end{figure} 

We show in Fig.~\ref{fig-5} the drift time distribution of the average pulse 
height summed over the adjacent pads, $\langle PH \rangle$, for pions and
electrons in case of a fibre radiator with 17~$\mu$m fibre diameter
(X=0.3 g/cm$^2$).
For electrons (square symbols) there is a significant increase 
in the average pulse height at higher drift times, due
to preferential absorption of TR at the entrance of the DC.
Note that for pions (circles) the average pulse height in 
the drift region exhibits a slight increase as function of DC's depth. 
This is the result of build-up of detector currents from ion tails,
convoluted in addition with the response of the preamplifier.
The peak at the beginning of these distributions originates from the primary
clusters in the amplification region, where the ionization is summed up in
the same time channel from both sides of the anode wires 
(see Fig.~\ref{fig-1}).
These features of the pion distribution have been reproduced by simulations
using GARFIELD \cite{rob}.
The dashed line in Fig.~\ref{fig-5} is the expected pulse height distribution
for electrons without TR; it has been obtained by scaling the pion 
distribution with a factor of 1.45, measured in a separate experiment without 
radiator.

Pulse height distributions as function of drift time were measured in
other experiments \cite{wat,zeus,hol,d0}, however, the observed trends were 
somewhat different.
A decrease of the pulse height as function of drift time was observed and 
it was attributed to electron attachment \cite{d0}.

\begin{figure}[hbt]
\centering\mbox{\epsfig{file=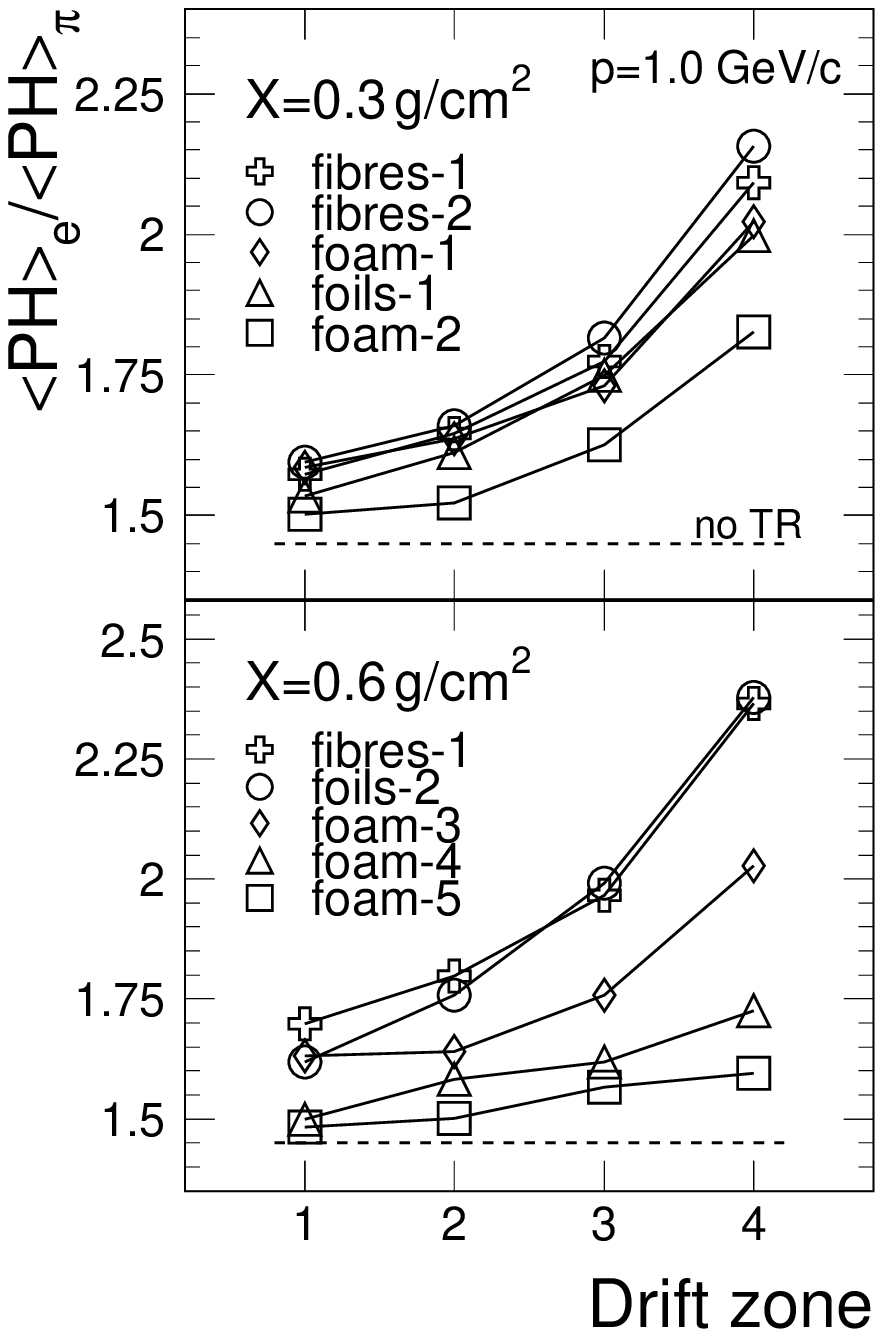, width=.44\textwidth}}
\caption{Average relative electron/pion pulse height as function of the 
drift zone for various radiators of two thicknesses (see Table~\ref{tab-1})
for the momentum of 1~GeV/c. Note the different scales on the vertical axis.}
\label{fig-6} \end{figure} 

To study the relative performance of the various radiators presented in 
Table~\ref{tab-1} we have classified them according to the equivalent 
thickness in two classes, with roughly X=0.3~g/cm$^2$ and X=0.6~g/cm$^2$ 
(for comparison, the radiation length for plexiglas is X$_0$=40.5~g/cm$^2$).
The compilation of measurements for the two classes is presented in
Fig.~\ref{fig-6} in terms of the ratio between the average pulse height
of electrons and pions, $\langle PH \rangle_e/\langle PH \rangle_\pi$,
as function of detector depth. 
As the final number of required samples is
yet to be specified by tracking performance considerations and due to 
statistics reasons as well, the detector depth is divided here in 4
drift zones, where each zone is a quarter of the chamber's drift region
(drift time between 0.75 and 2.35 $\mu$s in Fig.~\ref{fig-5}) and the 
numbering goes from the cathode wire plane towards the entrance of the DC.
In this representation, a better relative performance of the radiator
amounts to a higher ratio between electron and pion signal, while the
increase towards the entrance of the detector gives information about the 
characteristics of the spectrum of the TR.
The case of no TR would produce, for the present momentum value of 1~GeV/c, 
a flat distribution at about 1.45 (see above).

The most important conclusion from Fig.~\ref{fig-6} is that the fibre 
radiators exhibit performances comparable to that of radiators with foils.
Taking into account that the foil radiators  (with X=0.22~g/cm$^2$ and 
X=0.5~g/cm$^2$, respectively) are significantly lighter than the other 
radiators in both cases, our conclusion is in agreement with previous 
studies \cite{zeus}.
The fibres with lower density, {\it fibres-2}, produce slightly more TR compared 
to the more dense ones, {\it fibres-1}. In a separate study we have found that
the fibre diameter influences very little the TR yield. 
Radiators with fibres with 17 and 33~$\mu$m diameter show similar
TR performance for the same density and thickness (see Fig.~\ref{fig-7}
below).

Concerning the foams, their performance is comparable to the fibres
only in the case of the light PP foam, {\it foam-1}, however, with the 
disadvantage of a 10~cm thick radiator. The more packed version of the 
same material, {\it foam-2}, produces significantly less TR (furthermore, it is 
thicker, X=0.36~g/cm$^2$).
The two Rohacell foams exhibit very different features. 
Contrary to the expectations, it is the version with less structure
(invisible pores), {\it foam-3}, that gives higher TR yield.
The other Rohacell foam, {\it foam-4}, as well as the PE foam, {\it foam-5}, are
basically excluded as radiator candidates.
Judging by their apparent structure, these foams would have been expected 
to deliver reasonably good TR performance.
Their low TR yield may be the consequence of a higher absorption 
due to particular chemical compositions.
It was found that even PE foil radiators exhibit poor TR performance 
\cite{zeus}.
In general, similar results concerning the relative comparison of different 
radiator materials have been obtained in other experiments \cite{zeus,ed,wa89}.

A plexiglas 5.8~mm thick (X=0.66~g/cm$^2$) was used as dummy radiator 
and we have found that its contribution is essentially negligible. 
The results of the measurements for the dummy radiator and without radiator 
and the relative performance comparison of the 17~$\mu$m and 33 ~$\mu$m fibre 
radiators of identical thickness and density are summarized in 
Fig.~\ref{fig-7}.

\begin{figure}[hbt] 
\centering\mbox{\epsfig{file=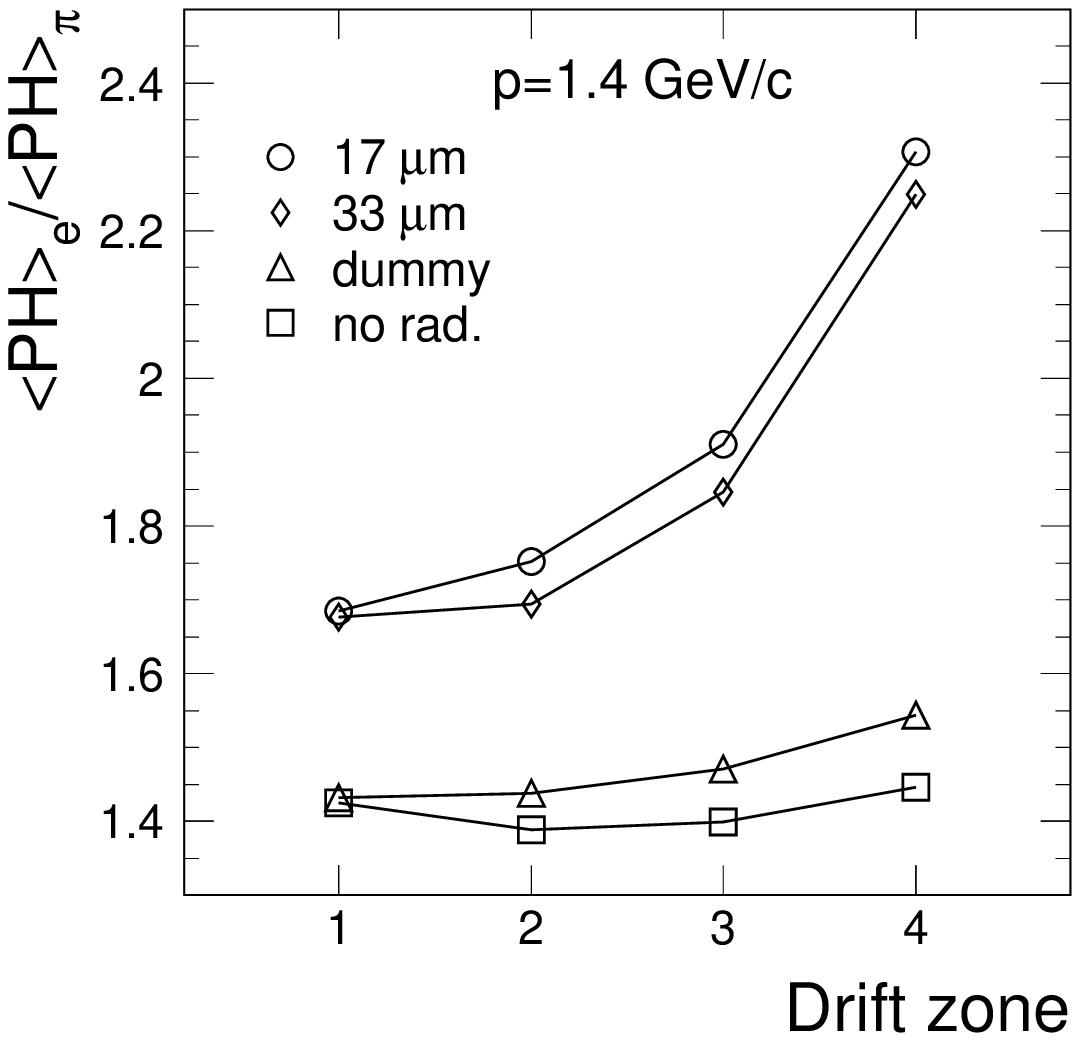, width=.4\textwidth}}
\caption{Average relative electron/pion pulse height as function of the 
drift zone for 17 and 33 ~$\mu$m fibre radiators for the momentum of 
1.4~GeV/c. Measurements without radiator and with a dummy one are shown
for comparison.}
\label{fig-7} \end{figure} 

The final choice of the radiator has to be a compromise between
TR yield, total thickness of material and, given the large size and the 
difficult collider geometry of the ALICE TRD, mechanical considerations. 
We have already tested sandwich radiators composed of fibres and foams and 
have found that their performance can satisfy the above requests.

\begin{figure}[htb]
\centering\mbox{\epsfig{file=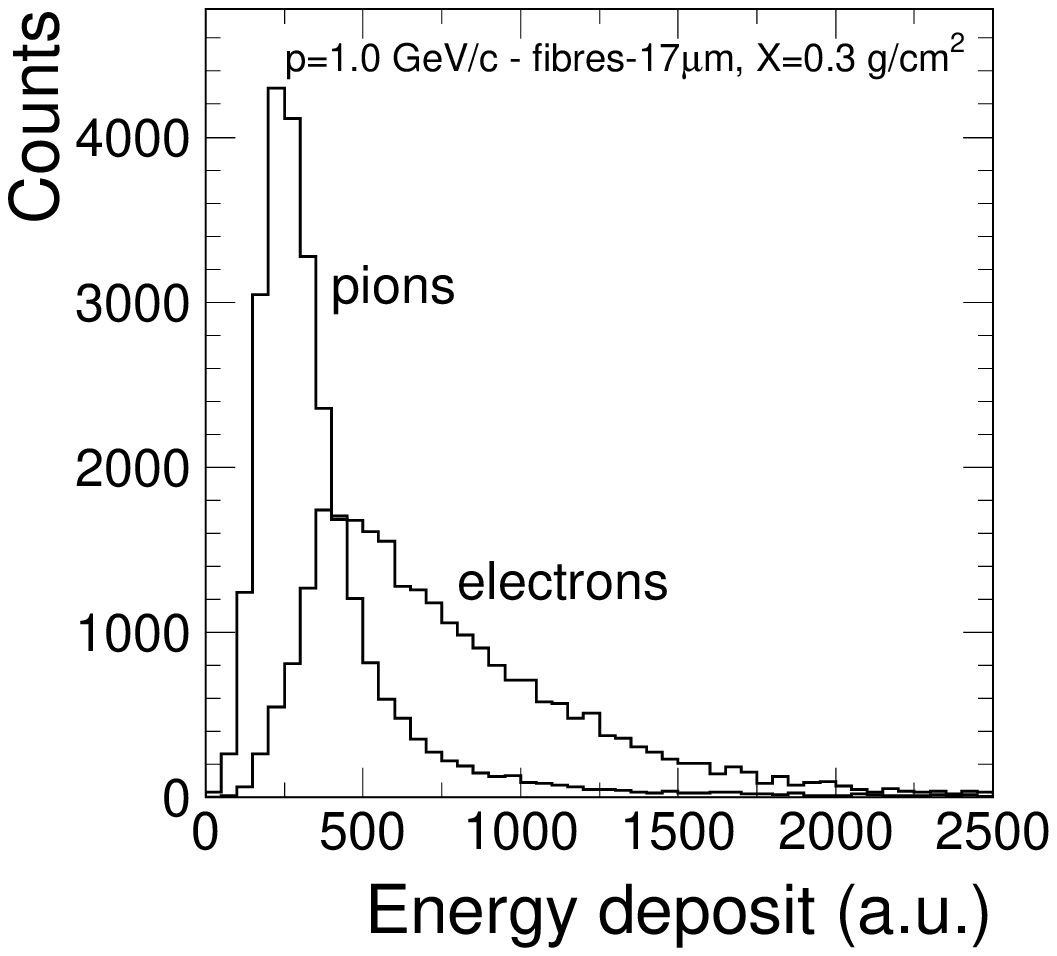, width=.45\textwidth}}
\caption{Integrated energy deposit for pions and electrons for
a momentum of 1.0~GeV/c. A radiator with 17~$\mu$m fibre has been used.}
\label{fig-8} \end{figure} 

The distributions of the integrated energy deposit are shown
in Fig.~\ref{fig-8} for pions and electrons for the momentum of 1~GeV/c, 
in case of a 17~$\mu$m fibres radiator.
The pure Landau distribution of pions is skewed towards higher values 
in the case of electrons by the contribution of the TR. 

\begin{figure}[hbt] 
\centering\mbox{\epsfig{file=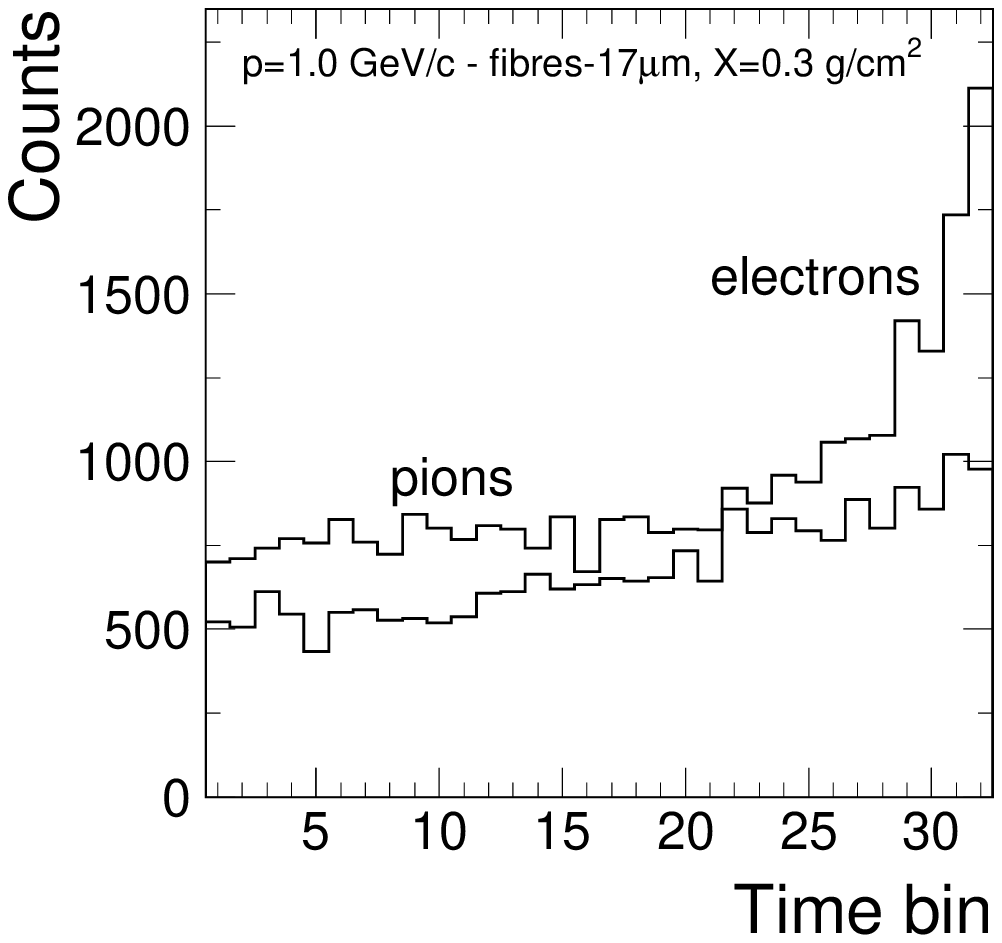, width=.42\textwidth}}
\caption{The distributions of position of the largest cluster found in the 
drift region for pions and electrons. The momentum is 1.0~GeV/c and the 
radiator is 17~$\mu$m fibres.}
\label{fig-9} \end{figure} 

The distributions of the position of the largest cluster 
found in the drift region are shown in Fig.~\ref{fig-9}.
The detector depth is expressed here in time bin (of 50 ns) number, where 
the counting starts at 0.75~$\mu$s (see Fig.~\ref{fig-5}) and 
increases towards the entrance window for a total of 32 time bins.
The trends seen in Fig.~\ref{fig-5} are present in these distributions as well.
For the case of electrons the probability to find the largest cluster is 
strongly increasing towards the entrance of the detector (higher time bin 
number) due mainly to the contribution of TR, while for pions there is only 
a slight increase which is due to the ion tail build-up explained above.

The distributions presented in Fig.~\ref{fig-8} and Fig.~\ref{fig-9} 
correspond to the same number of events.
These distributions have been employed as probability distributions in
simulations aimed at determining the pion rejection factor for the proposed
configuration of the ALICE TRD.
To extract the pion rejection factor we have studied three different methods:
i) truncated mean of integrated energy deposit - TMQ;
ii) likelihood on integrated energy deposit (see Fig.~\ref{fig-8}) - L-Q 
\cite{bun}; 
iii) bidimensional likelihood on energy deposit and position of the largest 
cluster found in the drift region of the DC (see Fig.~\ref{fig-9}) - L-QX
\cite{hol}.
We assume that the six layers have identical performance as represented by 
the measured distributions of Fig.~\ref{fig-8} and Fig.~\ref{fig-9}.
Both the truncated mean (the truncation is done by excluding the highest 
value of the integral energy deposit among the layers) and the 
likelihood (see i.e. \cite{bun,nom2} for details) distributions were 
constructed over the six (independent) layers for the same number of simulated 
pion and electron events.
Cuts of certain electron efficiency were involved at the end on these 
distributions and the pion efficiency is derived within these cuts.
We note that another method, ``cluster counting'' \cite{fab} is widely
used, in particular for ``fine grain'' TRDs like the ones used in ATLAS 
\cite{atlas} and in HERA-B \cite{herab}.
As was shown in \cite{zeus,hol} and as our own simulations have demonstrated
\cite{trd}, the likelihood on integrated charge remains better than the 
cluster counting method.

\begin{figure}[hbt]
\centering\mbox{\epsfig{file=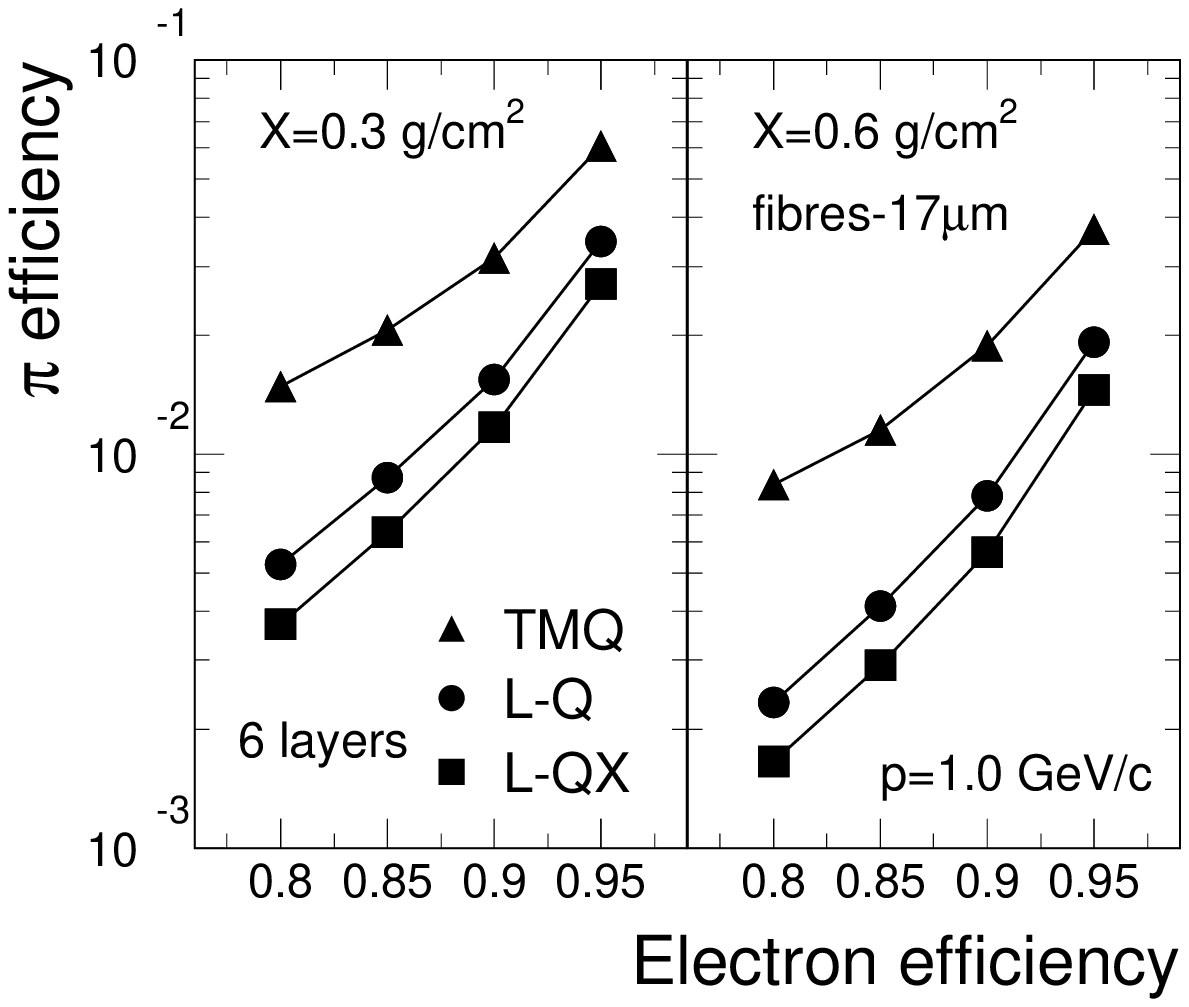, width=.48\textwidth}}
\caption{The pion efficiency as function of electron efficiency determined 
with truncated mean on energy deposit (TMQ), likelihood on total energy 
deposit (L-Q) bidimensional likelihood on charge deposit and DC depth (L-QX).}
\label{fig-10} 
\end{figure} 

In Fig.~\ref{fig-10} we present the pion efficiency (the inverse of the 
rejection factor) as function of electron efficiency (90\% electron 
efficiency is the commonly used value) in case of {\it fibres-1} radiators
for the momentum of 1~GeV/c. 
The three methods introduced above are compared.
The truncated mean method, although it delivers sizeably worse identification,
has the advantage of being very easy to use, being advantageous especially 
for an on-line identification.
The bidimensional likelihood delivers the best rejection factor and 
will be studied further in order to optimize the final detector design. 
As emphasized earlier \cite{hol}, the use of FADC to process the signals 
in a TRD can bring up to a factor of 2 in pion rejection power.
In general, the three methods employed here give results in good agreement 
with earlier studies \cite{zeus,hol}.

By doubling the equivalent thickness of the radiator from X=0.3~g/cm$^2$ 
(left panel of Fig.~\ref{fig-10}) to X=0.6~g/cm$^2$ (right panel) one gains 
a factor of about 2 in pion rejection power. However, as discussed before, 
it remains to be seen how the additional material will influence 
(by producing secondary particles)
the performance of the TRD itself and of other ALICE sub-detectors.

\begin{figure}[hbt]
\centering\mbox{\epsfig{file=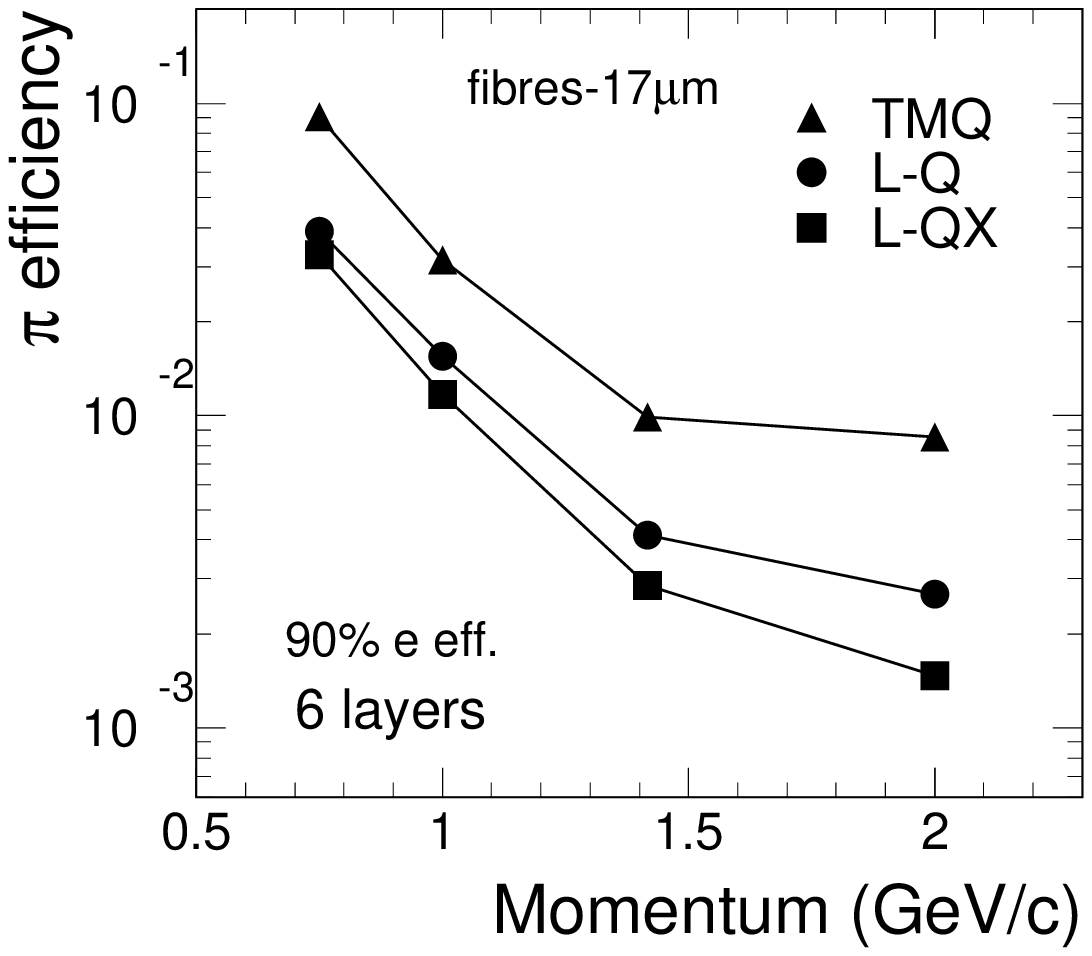, width=.48\textwidth}}
\caption{Pion efficiency as function of momentum for a radiator with 
17~$\mu$m fibres. The three methods used are discussed in the text.}
\label{fig-11}
\end{figure} 

The pion efficiency at 90\% electron efficiency as function of momentum 
is shown in Fig.~\ref{fig-11}.
The steep decrease of pion efficiency at momenta around 1~GeV/c is due to 
the onset of TR production \cite{bun,nom2}.
Towards our highest momentum value, 2~GeV/c, the pion efficiency reaches 
a saturation, determined by the TR yield saturation and by the pion 
relativistic rise. Due to these effects the pion rejection is expected to
get slightly worse for momenta above 3~GeV/c \cite{bun,wat,but,zeus}.
As one can see in Fig.~\ref{fig-11}, at momenta around 2~GeV/c the pion 
rejection factor of 300 to 500 achieved during these tests is above the 
required value for the ALICE TRD. However, one has to bear in mind that
a significant worsening of TRD performance has been registered when
going from prototype tests to real detectors \cite{dol}. This can be
the effect of detector loads in a multiparticle environment.
On the other hand, impressive pion rejection factors of 1000 and above 
have been achieved in full size TRDs
in the NOMAD \cite{nom2} and HERMES \cite{her} experiments.

\section{Conclusions}
We have presented results of test measurements with a prototype for the 
ALICE TRD.
By studying different radiators we have been able to select the best 
candidates for the final radiator, which has to be the best compromise 
between TR performance and mechanical stiffness.
Sandwich radiators composed of foam and fibres are currently under 
investigation.

The pion rejection capability was studied for different beam momenta
and for different methods of data analysis.
We have demonstrated that a rejection factor of about 500 (at an electron 
efficiency of 90\%) for momenta around 2~GeV/c can be achieved using a fibre 
radiator of up to 5~cm thickness.
Further work is underway, in particular devoted to the position resolution 
of the DC and to the performance of the detector in a high multiplicity 
environment.

\section*{Acknowledgment}
We acknowledge the great help of N.~Kurz on the data acquisition,
of J.~Stroth concerning the Si detectors and of M. Ardid for the assistance
in data taking and data analysis.

\nocite{*}
\bibliographystyle{IEEE}

\end{document}